\documentclass[sigconf]{acmart-arxiv}
\usepackage{booktabs}
\usepackage{multirow}
\usepackage{threeparttable}
\usepackage[linesnumbered,ruled,vlined]{algorithm2e}
\usepackage{listings, xcolor}

\newcommand{\ignore}[1]{}

\colorlet{punct}{red!60!black}
\definecolor{background}{RGB}{255,255,255}

\lstdefinelanguage{json}{
    basicstyle=\footnotesize \ttfamily,
    numbers=none,
    showstringspaces=false,
    breaklines=true,
    tabsize=2,
    frame=lines,
    backgroundcolor=\color{background},
    literate=
      {:}{{{\color{punct}{:}}}}{1} 
      {,}{{{\color{punct}{,}}}}{1},
}
\begin{document}

\title{Distant Supervision for Topic Classification\\ of Tweets in Curated Streams}

\author{Salman Mohammed, Nimesh Ghelani, and Jimmy Lin}
\affiliation{
  \institution{\vspace{0.1cm}David R. Cheriton School of Computer Science}
  \institution{University of Waterloo, Ontario, Canada}
}
\email{{salman.mohammed,nghelani,jimmylin}@uwaterloo.ca}

\begin{abstract}
We tackle the challenge of topic classification of tweets in the
context of analyzing a large collection of curated streams by news
outlets and other organizations to deliver relevant content to users.
Our approach is novel in applying distant supervision based on
semi-automatically identifying curated streams that are topically
focused (for example, on politics, entertainment, or sports).  These
streams provide a source of labeled data to train topic classifiers
that can then be applied to categorize tweets from more
topically-diffuse streams. Experiments on both noisy labels and human
ground-truth judgments demonstrate that our approach yields good
topic classifiers essentially ``for free'', and that topic
classifiers trained in this manner are able to dynamically adjust for
topic drift as news on Twitter evolves.
\end{abstract}

\maketitle

\vspace{-0.25cm}
\section{Introduction}

Our work tackles the problem of identifying interesting posts in
social media streams that are then delivered to users' mobile devices
in real time as push notifications. In our problem formulation, users
are interested in broad topics such as politics, technology, sports,
or entertainment, and we focus on tweets due to their widespread
availability. Although topic detection on Twitter (i.e., ``trends'')
is a well-trodden area, we take a novel approach:\ instead of trying
to tame the cacophony of unfiltered tweet streams, we exploit a
smaller, but still sizeable, collection of curated streams (accounts)
corresponding to different media outlets.

Our approach skirts many thorny issues in traditional approaches to
event detection, but requires solving two non-trivial
challenges. First, in order to obtain reasonable coverage of topics
and locales, we need to consider a volume of tweets that would still
be overwhelming for a user, and thus we need to identify tweets that
are salient and novel. Second, many streams contain tweets about a
multitude of topics, and therefore we need to develop topic
classifiers to separate posts into different categories.

This paper is focused on the second problem. We propose a novel
distant supervision technique for automatically gathering noisy topic
category annotations from topically-focused streams. These can be used
to train topic classifiers and applied to topically-diffuse streams to
retain only those tweets that a user might be interested in.
Experiments on both noisy labels and human ground-truth judgments
demonstrate that our approach yields good topic classifiers
essentially ``for free''. Experimental results show that having more
data and more recent data obtained through distant supervision
improves classification, and weighting training instances based on
recency yields additional gains.  Furthermore, our approach is able to
dynamically adapt to topic drift as news on Twitter evolves.

\section{Related Work}

There is already much work on Twitter event detection; see Atefeh and
Khreich~\cite{atefeh2015survey} for a recent survey. However, this
paper is primarily focused on topic classification of tweets.

Most work on tweet classification in the past has involved manual
judgments. Becker et al.~\shortcite{becker2011beyond} clustered tweets
in real time and performed binary classification between real events
and non-events. Kinsella et al.~\shortcite{kinsella2011topic} used
metadata from hyperlinked webpages to classify blog posts into different
topics. Lee et al.~\shortcite{lee2011twitter} worked on trending
topic detection by classifying tweets into 18 categories.  However,
obtaining manual annotations is costly and methods that depend on
them are likely to perform worse over time due to concept drift,
especially in the context of social media \cite{magdy2014adaptive}.

Distant supervision can overcome the challenges of obtaining manual
annotations and there has been previous work on using distant
supervision in topic classification. Husby and
Barbosa~\shortcite{husby2012topic} used Wikipedia articles labeled
with Freebase domains as training data to classify blog posts by
topics. Zubiaga et al.~\shortcite{zubiaga2013harnessing} used
human-edited web page directories to assign topic labels to tweets
that contained URLs to those pages. Magdy et
al.~\shortcite{magdy2015bridging,magdy2015distant} transferred labels
from YouTube videos to tweets that link to those videos. Our work
also takes advantage of distant supervision, but in a way that
directly leverages human curation ``for free''.

\section{Approach}

The starting point of nearly all event detection work on Twitter is an
unfiltered stream of tweets---the more tweets, the better. From this
cacophony, the system tries to identify events or ``trending''
topics. Such a needle-in-a-haystack approach is noisy and prone to
manipulation (fake news, ``astro-turfing''). Our work adopts a
completely different approach:\ we begin with the observation that
there already exist many human-curated streams of interesting events,
corresponding to the Twitter accounts of various media outlets. The
news editors at CNN, for example, tweet breaking news from @cnn
and related accounts. Almost every media outlet, large and small, has
their own Twitter account. We wonder, why not build event detection
techniques on a collection of these curated streams? Especially for
``head events'' of broad significance to large groups of users, such
an approach seems intuitive.

\begin{table*}[]
\centering
\caption{Example Twitter accounts illustrating {\it focused}, {\it hybrid}, and {\it general} streams, with example tweets.}
\label{tab:example-tweets}
\begin{tabular}{|c|c|c|}
\hline
\textbf{Account}                                                             & \textbf{Topic}                     & \textbf{Tweet}                                                                                                                                                \\ \hline\hline
\multirow{1}{*}{\begin{tabular}[c]{@{}c@{}}ESPN\\ (focused)\end{tabular}}   & sports                           & \begin{tabular}[c]{@{}c@{}}RT @SportsCenter: Cavs' Big 3 will not travel to Memphis with team\\ for Cleveland's 2nd night of home-road back-to-back vs Grizzlies...\end{tabular}  \\ \hline
\multirow{2}{*}{\begin{tabular}[c]{@{}c@{}}\\TheEconomist\\ (hybrid)\end{tabular}}  & politics                           & Vladimir Putin has capitalised on attempts  to discredit Western democracies...                                                                               \\ \cline{2-3} 
                                                                                 & business                           & \begin{tabular}[c]{@{}c@{}}Like other financial shares, the Dow has shot  up in the wake of Donald \\ Trump's election...\end{tabular}                        \\ \hline
\multirow{2}{*}{\begin{tabular}[c]{@{}c@{}}engadget\\ (hybrid)\end{tabular}}     & tech                               & Slack now has built-in video calling...                                                                                                                       \\ \cline{2-3} 
                                                                                 & gaming                              & `Pok\'emon Go' is live in India and South Asia...                                                                                                               \\ \hline
\multirow{3}{*}{\begin{tabular}[c]{@{}c@{}}CNN\\ (general)\end{tabular}}         & politics                           & \begin{tabular}[c]{@{}c@{}}Philippines President Duterte admits he  didn't attend sessions at a \\ summit in Laos to avoid US President Obama...\end{tabular} \\ \cline{2-3} 
                                                                                 & \multicolumn{1}{l|}{entertainment} & Mick Jagger's family tree has gotten more complicated...                                                                                                      \\ \cline{2-3} 
                                                                                 & tech                               & \begin{tabular}[c]{@{}c@{}}A gorilla emoji is part of Apple's latest \\ emoji  update. Bet you can guess what people are calling.…\end{tabular}               \\ \hline
\end{tabular}
\end{table*}

Our approach skirts many thorny issues in event detection, such as the
definition of an event, which has been the subject of much debate
dating back over a decade~\cite{Allan_2002}. To us, an event is simply
what the editors of the underlying curated streams deem
interesting. Our problem formulation does indeed simplify the event
detection problem, but two unresolved challenges remain:

First, although our techniques operate on curated streams of posts,
the combined volume of these streams is still beyond what any human
can consume. In our experiments, we observed around 16,000 tweets per
day on average over a period of 21 days from our curated streams.
Obviously, it is not possible for a user to consume all these
tweets. Furthermore, there are many duplicate tweets corresponding to
reports by different outlets. Thus, even over curated streams, we must
still identify what the {\it salient} and {\it novel} tweets are.

Second, curated streams vary in their topical focus. Some accounts
have a narrow focus, e.g., only entertainment news, while others have
broad coverage, i.e., they contain tweets about multiple topics.
Since users are often only interested in particular topics, we need
topic classifiers that can identify relevant tweets. This paper
focuses on this second problem, taking advantage of distant supervision
techniques to automatically acquire up-to-date training labels in real
time.

\begin{figure}[t]
	\centering
    \includegraphics*[scale=0.55]{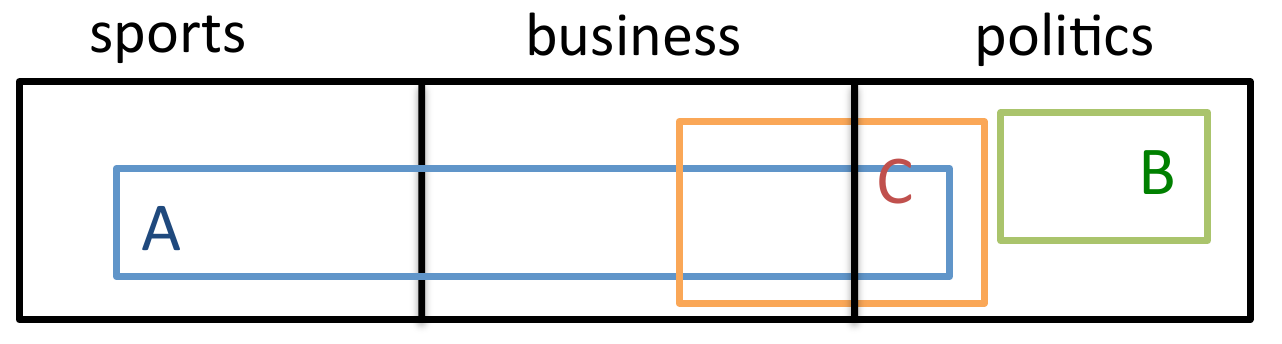}
    \caption{An illustration of the different topic streams.}
    \label{fig:topic-grid}
\end{figure}

Our problem can be schematically illustrated in
Figure~\ref{fig:topic-grid}. For simplicity, we only show three
topics. We illustrate streams in terms of the topic combinations that
they cover (here, they are streams A, B, and C). Streams can be
classified into three types:\ \textit{general}, which tweet about a
broad spectrum of news, \textit{focused}, which tweet about a very
specific topic, and \textit{hybrid}, which tweet about a few topics
but are not focused (see examples in Table~\ref{tab:example-tweets}). 
Our intuition is as follows:\ for event
detection, we would benefit from broad coverage tweets for additional
signal, but we must develop topic classifiers to discard tweets that
a particular user would not be interested in. We can exploit tweets
from \textit{focused} accounts to train topic classifiers using
distant supervision, which can then be used to classify tweets from
\textit{general} and \textit{hybrid} accounts---thus maximizing both
coverage as well as relevance. 

As a specific example, we might imagine that our user is interested in
receiving notifications about politics. We can build classifiers using
tweets from accounts with a narrow focus, e.g., ``politics'' (stream
B).  We can keep the relevant tweets from \textit{hybrid} and
\textit{general} accounts (streams A and C) by checking the predicted
labels from the politics classifier.  Note that our idea for using
focused accounts as distantly-supervised labels applies to both topics
as well as locales (e.g., US vs.\ UK), but we only focus on topic
classification in this paper.

\section{Data Collection}

Facebook published an article in May 2016 providing an overview of
their Trending Topics algorithm \cite{FacebookTrending}. The article
provided a list of RSS URLs, mapped to countries and topics, that
their algorithm uses to identify breaking events. Most of those URLs
correspond to popular news media accounts such as CNN and
ESPN. Although the Facebook data contain RSS feeds in many
languages, in this work we only focus on English feeds. Based on a few
simple heuristics and manual verification, we obtained a list of 293
Twitter accounts that correspond to media outlets in the original
Facebook dataset. We then manually classified each account as {\it
  focused}, {\it hybrid}, or {\it general} (based on the previous
section).

We monitored tweets from these 293 curated streams from December 13,
2016 00:00 UTC to January 3, 2017 9:11 UTC and received a total of
337,307 tweets.  Table~\ref{tab:twitter-accounts} shows the number of
Twitter accounts mapped to each topic and the number of tweets
observed. Note that counts for each topic do not sum up to the
\textit{total} row because a Twitter account can be mapped to more
than one topic (e.g., the {\it hybrid} accounts). For the purposes of
training, the {\it general} accounts are not helpful since they do not
provide training labels, although we do select tweets from them for
our manual evaluation (more details below). In our experiments, we did
not consider the category ``gaming'' due to the low prevalence of
tweets during our observation period.

For the experimental condition that we call ``noisy labels'',
we divided the collected data into an 80/20 training/test split
chronologically, using older data for training and newer data for
testing. For training, we used tweets from single-topic (i.e., {\it
  focused}) streams as positive examples, and randomly sampled negative examples from other
accounts to include up to five times the number of positive examples.

To create the ground truth, we performed manual judgments on tweets
pooled from 31 December, 2016 to 2 January, 2017. The pool was created
by randomly sampling no more than 50 tweets from randomly selected
\textit{hybrid} and \textit{general} accounts. We ensured that, in
total, there were more than 500 tweets each from \textit{hybrid} and
\textit{general} accounts. Since for many of the tweets we were not
able to assign one of the existing topic labels (these we labeled as
``other''), we continued pooling and assessing until we had at least
1000 labeled tweets for the topic labels we were interested in.
Table~\ref{tab:twitter-accounts} shows the distribution of labels in
the gold standard test set. Note that in our annotation process we
allowed a tweet to be assigned multiple topic labels, hence the
individual counts do not sum up to the total. In the experiments using the gold
standard test set, the training set consists of tweets till 30 December, 2016.

\begin{table}[]
\centering
\caption{Summary of the topics: number of accounts and composition of the test sets.}
\label{tab:twitter-accounts}
\begin{tabular}{|c|c|c|c|}
\hline
\multirow{2}{*}{\textbf{Topic}} & \multirow{2}{*}{\textbf{Accounts}} & \multicolumn{2}{|c|}{\textbf{Tweets in Test Set}} \\ \cline{3-4}
&  & \textbf{Noisy} & \textbf{Gold}  \\ \hline\hline
general        & 68  & - & -           \\ \hline
politics       & 33  & 5641 & 346            \\ \hline
business       & 23  & 2401 & 69            \\ \hline
health         & 40  & 4031 & 65            \\ \hline
sports         & 30  & 9290 & 73            \\ \hline
science        & 25  & 947  & 56           \\ \hline
technology     & 35  & 2616 & 130           \\ \hline
entertainment  & 50  & 8077 & 332           \\ \hline
gaming         & 17  & 931  & -        \\ \hline
\hline
other        & -   & -    & 521        \\ \hline \hline
\textbf{total}          & \textbf{293} & \textbf{33934}& \textbf{1536}         \\ \hline
\end{tabular}
\end{table}

\section{Experimental Results}

Based on the data preparation process described above, for each topic,
we trained an individual classifier via distant supervision using
tweets from single-topic {\it focused} accounts. We extracted TF-IDF
features from the words in the tweets that were obtained from the NLTK
tweet tokenizer. The logistic regression classifier from sci-kit learn
was used with default parameters. In addition to the individual
classifiers, we also trained a multi-class classifier using the
multinomial na\"ive Bayes implementation in sci-kit learn.

In all our experiments, we used the test sets as described in the previous
section. Given a list of chronologically-ordered training examples $\langle
t_1,t_2,\ldots,t_N\rangle$, where $N$ is the total number of available training
examples, the training set ($\mathcal{D}_{\textrm{train}}$)
represents a continuous sublist $\langle t_i,t_{i+1},\ldots,t_{i+m}\vert i\ge
1, i+m\le N\rangle$.
We varied $\mathcal{D}_{\textrm{train}}$ in two sets of experiments:

In the first set of experiments, we fixed the end of
$\mathcal{D}_{\textrm{train}}$ to the latest available tweet ($t_N$), 
and varied the amount of training data by moving the start of
$\mathcal{D}_{\textrm{train}}$. More formally, $\mathcal{D}_{\textrm{train}}$ is
a sublist $\langle t_i,t_{i+1},\ldots,t_{N}\vert i\ge
1\rangle$. This lets us examine the effect of providing the
classifier more historical training data.

In the second set of experiments, we fixed the size of
$\mathcal{D}_{\textrm{train}}$ to $0.5 \cdot N$ for the evaluation by ``noisy
labels'' and to $0.6 \cdot N$ for the ground-truth judgments. Then we varied
the recency of the training data by moving the start and end of
$\mathcal{D}_{\textrm{train}}$. More formally, $\mathcal{D}_{\textrm{train}}$ is
a sublist $\langle t_i,t_{i+1},\ldots,t_{i+R{\cdot}N}\vert i\ge
1, i+R{\cdot}N\le N\rangle$, where $R=0.5$ or $R=0.6$. This lets us examine the impact of
training on data that is ``out of date'', thus quantifying the effects
of topic drift.

\begin{figure*}[t]
	\centering
    \includegraphics*[scale=0.40]{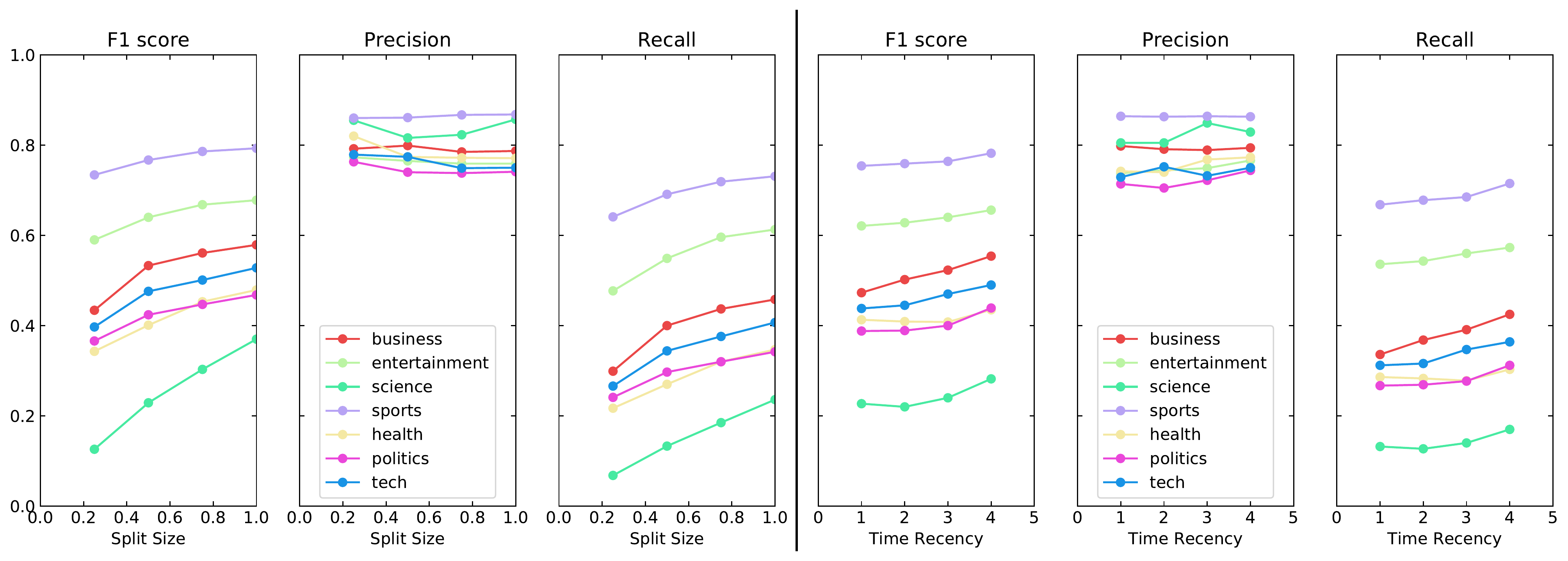}
\vspace{-0.25cm}
    \caption{Evaluation on noisy labels:\ the effect of using more training data (left) and using more recent training data (right).}
    \label{fig:silver-standard}
\end{figure*}

\begin{figure*}[t]
    \includegraphics*[scale=0.40]{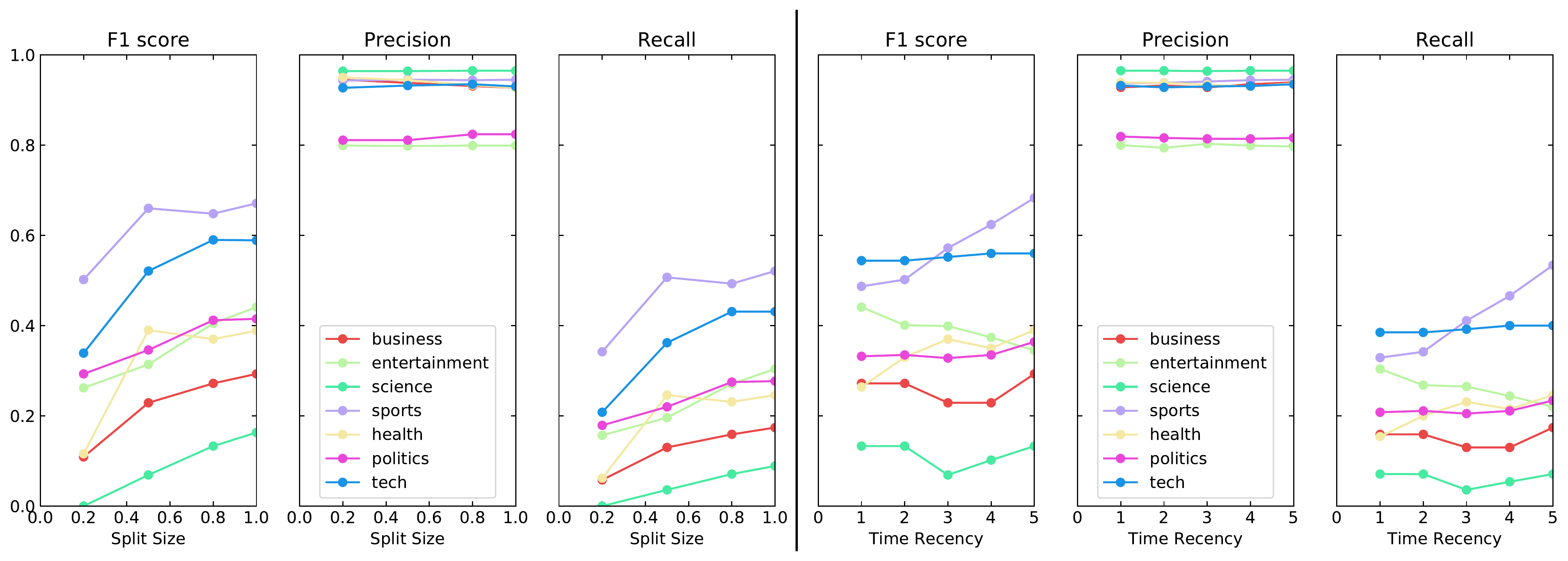}
\vspace{-0.25cm}
    \caption{Evaluation on human-annotated ground truth:\ the effect of using more training data (left) and using more recent training data (right).}
    \label{fig:gold-standard}
\end{figure*}

Figure~\ref{fig:silver-standard} and Figure~\ref{fig:gold-standard}
show the plots of the F1, precision, and recall scores for the
different topics on the noisy labels and human-labeled ground truth,
respectively. The plots show an upward trend when the split size
increases, i.e., more training data is used whose labels were obtained
through distant supervision. This result suggests that more data yield
better classification accuracy, which is not surprising.  The plots on
the right-hand side shows that training on recent time splits (keeping
the size constant) generally improves topic classification. This
suggests that classifier effectiveness suffers from topic drift, which
also makes sense since Twitter content reflects successive news
cycles.

Overall, evaluation results on both the noisy labels and human-labeled
ground truth are consistent and support the same conclusions.
Entertainment news shows a decrease in F1 score on the human judgments,
but this is likely due to the fact that many tweets from the
\textit{general} accounts---representing a wide variety of topics
such as music, travel, food, and film---were labeled as ``entertainment''
news. The results for ``science'' news are significantly worse than the
other topics due to its low prevalence (so there is little training
data to begin with).

\begin{table}[t]
\centering
\caption{Differences in F1 scores when the samples were weighted ($p=10$) during classification.}
\label{tab:delta-f1-weghting}
\begin{tabular}{|c|c|c|}
\hline
\textbf{Topic} & \textbf{$\Delta $ F1 - (Noisy)} & \textbf{$\Delta $ F1 - (Truth)} \\ \hline\hline
business       & 0.09                     & 0.13                   \\ \hline
entertainment  & 0.03                     & 0.07                   \\ \hline
science        & 0.10                     & 0.14                   \\ \hline
sports         & 0.02                     & 0.05                   \\ \hline
health         & 0.06                     & 0.16                   \\ \hline
politics       & 0.05                     & 0.10                   \\ \hline
tech           & 0.05                     & 0.12                   \\ \hline \hline
all            & 0.07                     & 0.04                   \\ \hline
\end{tabular}
\vspace{-0.25cm}
\end{table}

In attempt to incorporate advantages of recent data along with more
data, we tried weighting training examples based on their
recency. Given a list of chronologically-ordered tweets $\langle
t_0,t_1,\ldots,t_n\rangle$, we weight tweet $t_i$ as $e^{(\log(p)
  \cdot i)/n}$.  In other words, we sampled weights using an
exponential function such that weights for $t_0$ and $t_n$ are $1$ and
$p$, respectively. When $p$ is set to $1$, the training is equivalent
to unweighted training. We observed empirically that classifier
effectiveness improves as we increase $p$ up to $50.0$. The rate of
improvement decreases as $p$ increases, and is expected to decrease
for higher values of $p$. However, our experimental results are
reported using $p=10$.

We trained the logistic regression classifier with
and without weighting the training samples, as described above.
For evaluation with the noisy labels, we used the same 80/20 training/test
split as in the previous experiments. For evaluation using the ground truth labels,
we trained on all tweets up until 30 December, 2016 and tested on
all available manual judgments.
Results in Table~\ref{tab:delta-f1-weghting} show large improvements in F1
scores when our weighting scheme is used in place of uniform weighting.

\section{Conclusion}

In this paper we tackle the problem of topic classification for tweets
in the context of pushing useful notifications to users interested in
broad topic categories. We use distant supervision to obtain topic
labels by identifying Twitter accounts with a narrow focus, the
contents of which serves as training data for logistic regression
classifiers. Experimental results show that classifier effectiveness
improves with more data and also with more recent data. Weighting
recent samples yields further improvements, and results suggest that
there are noticeable topic drift effects, but that our techniques are
able to compensate. Overall, we empirically demonstrate the
effectiveness of a novel approach to gathering topic labels for
tweets, practically ``for free''. We are in the process of building a
system that leverages these techniques and are planning to conduct
field studies involving real users.

\section{Acknowledgments} 

This work was supported in part by the Natural Sciences and
Engineering Research Council (NSERC) of Canada.


\end{document}